\documentclass[%
 reprint,
nofootinbib,
 amsmath,amssymb,
 aps,
]{revtex4-2}

\usepackage{bm}% bold math
\usepackage{cleveref}
\usepackage{dcolumn}% Align table columns on decimal point
\usepackage{textcomp}
\usepackage{gensymb}
\usepackage{graphicx}% Include figure files
\usepackage{physics}

\usepackage{changepage}   % for the adjustwidth environment

\newcommand{\deltaL}{\delta L}

\newcommand{\Eref}{E_0}
\newcommand{\Lref}{L_0}
\newcommand{\deltaLref}{\delta L_{0}}
\newcommand{\deltavref}{\delta v_{0}}

\newcommand{\Mplanck}{M_{\rm{Planck}}}
\newcommand{\Lplanck}{L_{\rm{Planck}}}
\newcommand{\deltaLplanck}{\delta L_{\rm{Planck}}}

\newcommand{\Lcoh}{L_{\rm{coh}}}
\newcommand{\Drho}{\mathcal{D}[\rho]}

\begin{document}

\preprint{APS/123-QED}

\title{Neutrino signals of lightcone fluctuations resulting from fluctuating space-time}

% Force line breaks with \\

\author{Thomas Stuttard}
\affiliation{%
Niels Bohr Institute, University of Copenhagen, DK-2100 Copenhagen, Denmark\
}%

\date{\today}

\begin{abstract}

One of the most common expectations of a quantum theory of gravity is that space-time is uncertain or fluctuating at microscopic scales, making it a stochastic medium for particle propagation. Particles traversing this space-time may experience fluctuations in travel times or velocities, together referred to as lightcone fluctuations, with even very small effects potentially accumulating into observable signals over large distances. In this work we present a heuristic model of lightcone fluctuations and study the resulting modifications to neutrino propagation, including neutrino decoherence and arrival time spread. We show the expected scale of such effects due to `natural' Planck scale physics and consider how they may be observed in neutrino detectors, and compare the potential of neutrinos to $\gamma$-ray astronomy. Using simulations of neutrino mass states propagating in a fluctuating environment, we determine an analytic decoherence operator in the framework of open quantum systems to quantitatively evaluate neutrino decoherence resulting from lightcone fluctuations, allowing experimental constraints on neutrino decoherence to be connected to Planck scale fluctuations in space-time and $\gamma$-ray results.

\end{abstract}

\maketitle

%\tableofcontents

\section{Introduction}

The quantum nature of gravity is one of the greatest open questions in fundamental physics, and despite decades of effort no complete theory of quantum gravity has been forthcoming. A major obstacle in this endeavour has been the paucity of experimental data constraining potential quantum gravitational effects in any meaningful way, which is a consequence of the incredible weakness of gravity compared to the other known fundamental forces. Indeed, significant quantum gravity effects are generally only expected at the Planck scale, meaning extremely high energies ($E \sim \Mplanck \sim 1.2 \times 10^{19}$ GeV, e.g. the \textit{Planck mass}), or small distances ($L \sim \Lplanck \sim 1.6 \times 10^{-35}$ m, e.g. the \textit{Planck length}).

In recent years however, experimental constraints on potential Planck scale quantum gravity effects have been achieved using high-energy particles of cosmological origin, exploiting observations of photons from distant gamma ray bursts (GRBs),  quasars and quiescent gas clouds~\cite{Lieu:2003ee, Abdo2009, HESS:2011aa, Perlman_2015, Vasileiou2015, PhysRevD.99.083009, Cooke:2020rco}, and the high-energy astrophysical neutrinos observed by neutrino telescopes~\cite{AMELINOCAMELIA2016318, ELLIS2019352, PhysRevD.102.063027, Wei:2018ajw} such as the IceCube neutrino observatory~\cite{Aartsen:2016nxy}. These measurements have achieved sensitivity to very weak effects due to the vast distances traversed by the observed particles, potentially allowing even weak effects to accumulate into measurable signals.

In the absence of an accepted model of quantum gravity, heuristic models of the potential characteristics or effects of quantum gravity are often invoked in experimental searches. A common expectation of quantum gravity is that the structure of space-time itself could be subject to the uncertainty principal and fluctuate at very small distance scales~\cite{PhysRev.97.511, misner1973gravitation}. For instance, the very geometry or curvature of space-time may fluctuate, in turn introducing intrinsic uncertainty/fluctuations in defining distance and time.  Additionally, it has been conjectured that the fluctuating nature of space-time could manifest as \textit{virtual black holes} (VBH)~\cite{Hawking:1995ag, tHooft:2018waj}, the quantum gravitational analogue of the virtual electron-positron pairs in the well known phenomenon of \textit{vacuum polarisation} in quantum electrodyamics (QED). This uncertain/fluctuating space-time is variously referred to as \textit{space-time foam}, \textit{quantum foam} or \textit{fuzzy space-time}~\cite{Hawking, PhysRev.97.511}.

A direct consequence of these space-time fluctuations are so-called \textit{lightcone fluctuations}, e.g. an intrinsic variability in the travel distance/time -- or indeed velocity -- for a particle propagating through this fluctuating space-time~\cite{PauliLightcone, RevModPhys.29.417, PhysRevLett.13.114, Ford_1995}. This variability can in principal produce measurable signals, such as a variability in arrival times of particles from distant sources such as GRBs~\cite{AMELINOCAMELIA2016318, Vasileiou2015}, and interference effects between otherwise coherent wave-like phenomena such as image degradation in $\gamma$-ray astronomy~\cite{Lieu:2003ee, Perlman_2015} or neutrino flavour decoherence~\cite{Hawking, Ellis:1983jz, Mavromatos2005, Anchordoqui:2005gj}. Lightcone fluctuation effects have been proposed in the contexts of D-brane recoils~\cite{Ellis:1999jf}, compactified space-times~\cite{Yu_2009}, gravitons~\cite{Ford_1995} and loop quantum gravity~\cite{PhysRevD.59.124021}. 

Searches for signatures of lightcone fluctuations offer one of only a handful (and arguably the most model independent) avenues to experimentally probe quantum gravity. To date, constraints on lightcone fluctuations resulting from fluctuating space-time largely derive from astrophysical photon observations. Neutrino signals however are less well explored, and offer a number of advantages over other cosmic messenger particles. The feeble interactions between neutrinos and matter allow them to travel vast distances completely unhindered, unlike photons for which the Universe is opaque at high energies. Additionally (and relatedly), astrophysical neutrinos are observed at energies far in excess of cosmological photon sources, reaching PeV and potentially EeV energies (compared to TeV for photons).

In a previous work~\cite{PhysRevD.102.115003} we investigated quantum gravity signals resulting from neutrino interactions with VBHs, demonstrating sensitivity to Planck scale physics is achievable with atmospheric neutrinos (travelling terrestrial baselines). Here we instead investigate neutrino signals from lightcone fluctuations in a heuristic model of fluctuating space-time, including arrival time spread and neutrino decoherence, and consider the expected size of these signals from `natural' Planck scale physics and their detection prospects. In particular, we for the first time evaluate the impact of travel distance uncertainty models employed in $\gamma$-ray quantum gravity searches on neutrino flavour measurements, determining an operator representing decoherence from lightcone fluctuations in the formalism of open quantum systems. This allows  experimental constraints on neutrino decoherence to be interpreted with respect to underlying Planck scale fluctuations, and directly compared to $\gamma$-ray results.

\section{Lightcone fluctuations}
\label{sec:lightcone_fluctuations}

Here we present a heuristic model of lightcone fluctuations, specifically of the accumulated uncertainty in a particle's travel distance as a function of distance and particle energy. 

The fundamental parameter of this model is the distance uncertainty, $\deltaLref$, associated with a particle travelling a reference distance, $\Lref$. The accumulation of this uncertainty over a distance $L$ is expressed as:

\begin{equation}
\delta L(L) = \deltaLref(L) \left( \frac {L} {\Lref} \right)^m ,
\label{eq:deltaL_no_E_dep}
\end{equation}

\noindent where the distance dependence is assumed to follow a power-law characterised by the index $m$, which is a free parameter of the model. $m$ can be predicted for a given concrete fluctuating space-time model, or instead can be fitted to data. Interpretation of the value of $m$ is discussed in Section \ref{sec:distance_dependence}.

We additionally consider the possibility that this distance uncertainty has a dependence on the particle's energy, given that Planck scale physics is commonly expected to be suppressed at energies below $\Mplanck$. An intuitive picture of this is that lower energy particles are less able to resolve the microscopic fluctuating nature of space-time. We therefore modify \Cref{eq:deltaL_no_E_dep} to include a power-law energy dependence characterised by the index $n$, which like $m$ can be either predicted or fit to data, and a reference energy scale, $\Eref$:

\begin{equation}
\delta L(E, L) = \deltaLref \left( \frac {L} {\Lref} \right)^m \left( \frac {E} {\Eref} \right)^n .
\label{eq:deltaL}
\end{equation}

Similar phenomenological forms for the energy dependence of Planck scale physics have been assumed in neutrino decoherence searches~\cite{PhysRevLett.85.1166, Anchordoqui:2005gj, Coloma:2018idr, PhysRevD.102.115003}. 

When considering Planck scale physics, a `natural' choice of reference values is $\Eref = \Mplanck$ and $\Lref = \Lplanck$, yielding:

\begin{equation}
\delta L(E, L) = \deltaLplanck \left( \frac {L} {\Lplanck} \right)^m \left( \frac {E} {\Mplanck} \right)^n ,
\label{eq:deltaL_planck}
\end{equation}

\noindent where $\deltaLplanck$ then represents the uncertainty in travelling one Planck length. This parameter can be fit to experimental data, and given that the Planck length is expected to represent the smallest measurable distance in Nature, a `natural' expectation would be:

\begin{equation}
\deltaLplanck = \Lplanck
\label{eq:deltaLplanck_natural}
\end{equation}

which leads to the following `natural' distance-uncertainty expression: 

\begin{equation}
\delta L(E, L) = \Lplanck^{1-m} L^m \left( \frac {E} {\Mplanck} \right)^n .
\label{eq:deltaL_natural}
\end{equation}

In the absence of energy dependence (e.g. $n=0$), \Cref{eq:deltaL_natural} reduces to a form used in a number of previous works~\cite{Ng:2003ag, Ng:2004qq, Perlman_2015, Cooke:2020rco}, characterised by single free parameter $\alpha$, referred to as the \textit{accumulation parameter}, which is the equivalent of $m$ in this work (where $\alpha = 1 - m$). 

We caution that the energy scale of quantum gravity may differ from $\Mplanck$, and therefore experimental searches should keep an open mind as to the value of $\Eref$.

\subsection{Interpretation of distance dependence}
\label{sec:distance_dependence}

The distance dependence defined in \Cref{eq:deltaL} can be characterised as~\cite{Ng:2004qq}: 

\begin{itemize}
    \item $m = 0$: The distance uncertainty has no distance dependence, e.g. does not accumulate. This implies either that the uncertainty is fundamentally distance independent (e.g. one might consider that the Planck length is the fundamental measurement precision limit of the Universe, regardless of the actual distance being measured), or that the fluctuations experienced by the particle as it travels are fully anti-correlated and cancel.
    \item $m = 1/2$: The distance uncertainty accumulates as $\delta L(L) \propto L^{1/2}$, which is characteristic of the accumulation of uncorrelated fluctuations (the so-called \textit{random walk} model).
    \item $m = 1$: The distance uncertainty accumulates as $\delta L(L) \propto L$. Such a scenario is expected if the fluctuations experienced by the particle are fully-correlated. 
\end{itemize}

The cases $m = 0$ and $m = 1$ are therefore the bounding cases, representing the most pessimistic and optimistic scenarios respectively. The case $m=1/2$ can be considered a relatively `natural' scenario, implying that fluctuations in one region of space are independent of those in another spatially separated region. A mildly anti-correlated scenario consistent with the \textit{holographic principal} is given by $m=1/3$~\cite{Ng:2003jk, Perlman_2015}. The $m = 1/2$ and $m = 1$ scenarios are explicitly tested via Monte Carlo (MC) simulations in Section \ref{sec:decoh_mc}.

\subsection{Natural expectation for distance uncertainty}
\label{sec:distance_uncertainty_natural}

\begin{figure}[htp]
    \centering
    \includegraphics[trim=0.5cm 0.0cm 0.0cm 0.0cm, clip=true, width=\linewidth]{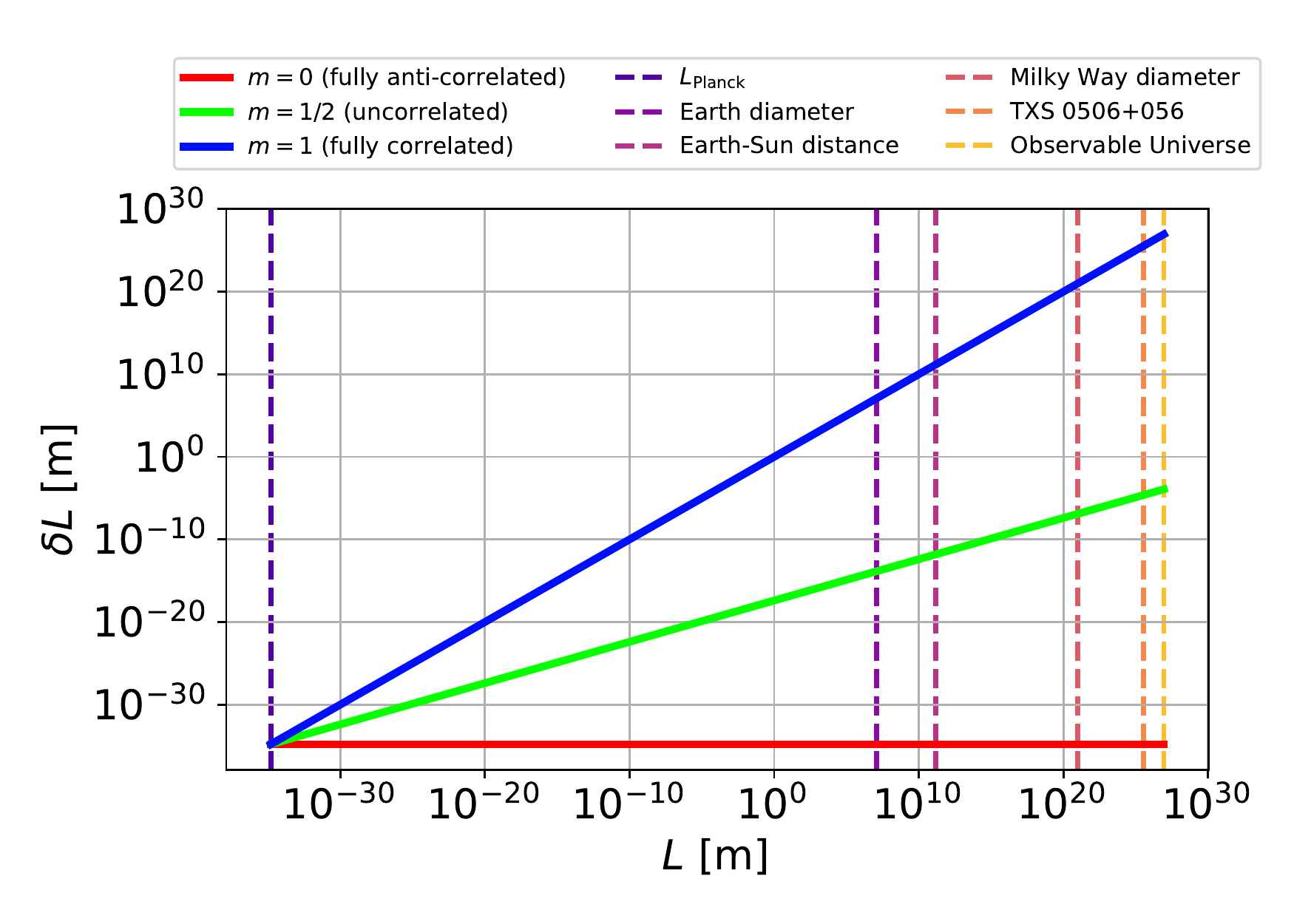}
    \caption{Distance uncertainty expected for the `natural' scenario given by \Cref{eq:deltaL_natural} for a particle with $E=\Mplanck$ (or $n=0$), as a function of particle travel distance. A number of reference distances are shown with dashed lines. Scenarios with differing $m$ are shown, with their interpretations discussed in Section \ref{sec:distance_dependence}.}
    \label{fig:delta_L_vs_L}
\end{figure}

\Cref{fig:delta_L_vs_L} shows the accumulation of distance uncertainty for the `natural' expectation defined by \Cref{eq:deltaL_natural}, for a range of different accumulation scenarios (e.g. $m$ values). The assumed particle energy is $E=\Mplanck$, e.g. this is a maximal bounding case, or alternatively represents particles of any energy if an energy independent scenario (e.g. $n = 0$) is assumed.

The distance uncertainty accumulated over large distances varies greatly depending on $m$. The most pessimistic case, $m=0$, yields an uncertainty of $\Lplanck$ regardless of distance, which is essentially unmeasurable. On the other extreme, the optimistic scenario, $m=1$, results in $\deltaL \sim L$, which is only viable if such effects are suppressed at energies below the Planck scale (e.g. $n>0$).

For the more `natural' uncorrelated $n=1/2$ case, the distance uncertainty accumulated over cosmological distances is $\order{ \rm{\micro m - mm} }$, even for a particle with Planckian energy. This effect, although small, is potentially feasible to study. However, even a weak suppression with energy would render these effects unmeasurable at the particle energies we are able to observe.

\subsection{Velocity fluctuations}
\label{sec:velocity_fluctuations}

In addition to fluctuations in travel distance, a related form of lightcone fluctuation that has been considered in the context of fluctuating space-time is velocity fluctuations~\cite{Vasileiou2015}, also referred to as \textit{stochastic Lorentz invariance violation}, which would result from any stochastic modifications to a particle's dispersion relation. Such a scenario is phenomenologically similar to distance fluctuations, as both result in fluctuations to a particle's travel time between two points. However, in the case of distance fluctuations, the particle's velocity remains unchanged from the particle's own perspective, whereas an observer sees an apparent fluctuation in velocity due to the fluctuating distance. The inverse is true when velocity fluctuations are the underlying mechanism. 

A phenomenological form for velocity fluctuations proposed in \cite{Vasileiou2015} is:

\begin{equation}
\delta v = \deltavref \left( \frac {E} {\Eref} \right)^n ,
\label{eq:delta_v}
\end{equation}

\noindent where $\deltavref$ represents velocity fluctuation for a particle with energy $\Eref$, with the energy dependence characterised in a similar manner to the distance fluctuations in \Cref{eq:deltaL}.

From standard uncertainty propagation we see that apparent velocity fluctuations resulting from underlying distance fluctuations are given by:

\begin{equation}
\delta v = v \frac{\delta L}{L} ,
\label{eq:delta_v_vs_delta_L}
\end{equation}

\noindent which combined with \Cref{eq:deltaL} gives:

\begin{equation}
\delta v = v \frac{\deltaLref L^{m-1}}{\Lref^m} \left( \frac {E} {\Eref} \right)^n .
\label{eq:delta_v_vs_L}
\end{equation}

The distance independent velocity fluctuation expression in \Cref{eq:delta_v} is recovered from \Cref{eq:delta_v_vs_L} when $m=1$ and $\deltaLref = \Lref$, implying $\deltavref = v$ in this case. This is consistent with the proposed natural scenario $\delta v(\Mplanck) = c$ proposed in \cite{Vasileiou2015}. 

We therefore see that velocity fluctuations can also be represented in terms of the distance fluctuation model proposed in this work (e.g. \Cref{eq:deltaL}), even if distance fluctuations are not the underlying mechanism, and thus experimental constraints on the parameters of this model constrain both velocity and distance fluctuation scenarios.

Note that an implication of this is that the distance independent velocity fluctuations of the form in \Cref{eq:delta_v} can only be the result of distance fluctuations with $m=1$, e.g. $\deltaL \propto L$. As discussed in Section \ref{sec:distance_dependence}, this corresponds to the highly optimistic fully-correlated distance fluctuation scenario. Distance independent velocity fluctuations are therefore unlikely to result from underlying distance fluctuations, and thus some other underlying fuzzy space-time mechanism is likely required to explain such a phenomenon, such as interactions with the virtual black hole or string/brane backgrounds. Such scenarios can be constrained by placing experimental constraints on $\deltavref$ and/or $\Eref$.

\section{Neutrino signals of lightcone fluctuations}
\label{sec:neutrino_signals}

We now consider the influence of the lightcone fluctuations described in Section \ref{sec:lightcone_fluctuations} on neutrino propagation, and the potential observable signals that could result. We explore two possibilities here; neutrino decoherence and arrival time fluctuations.

\subsection{Neutrino decoherence}

One of the major consequences of lightcone fluctuations is the loss of coherence of wave-like phenomena due to the variability in particle propagation distances/times, resulting in the potentially detectable degradation of superposition phenomena. 

One of the key proposed observable consequences of such effects is blurring/degradation of images in high energy photons from cosmological sources. For example, fluctuating photon propagation would degrade the wavefront at a telescope aperture, potentially preventing the formation of Airy disks~\cite{Lieu:2003ee, Perlman_2015}. More generally, lightcone fluctuations would blur photon point source images and ultimately render them undetectable once the fluctuations are comparable in scale to the photon wavelength~\cite{Perlman_2015}. Studies of these effects have enabled distance fluctuations to be constrained at the natural Planck scale for correlated, uncorrelated and even some anti-correlated scenarios, albeit only in cases where the effects are not suppressed by energy (e.g. $n=0$)~\cite{Perlman_2015}. Similar arguments also predict the degradation of a narrow FeII absorption line in photon spectra, with a recent study~\cite{Cooke:2020rco} also yielding a natural Planck scale constraint on such effects (again only for energy independent scenarios). 

Far less explored is the impact of the loss of coherence in neutrino propagation resulting from lightcone fluctuations in fluctuating space-time scenarios. A neutrino propagates as a superposition of three quantum states, known as \textit{mass eigenstates}. These are distinct from and misaligned with respect to the states in which the the neutrinos undergo interactions via the weak nuclear force, known as \textit{flavor eigenstates}. Together with the differing masses of the mass states, this produces the phenomena of \textit{neutrino oscillations}, whereby a neutrino produced in one flavor state may be detected as another. A neutrino therefore acts as a quantum interferometer, and is intrinsically sensitive to the fluctuations considered in this work. 

Neutrinos propagating through fluctuating space-time will become increasingly and stochastically out of phase with one another. This loss of coherence results in a damping of neutrino oscillations over distance, in a phenomenon known as \textit{neutrino decoherence}. Neutrino decoherence  has been the subject of a number of experimental searches which are often cited as sensitive to quantum gravity, but the connections of these measurements to potential underlying models (heuristic or otherwise) is little explored. In previous work~\cite{PhysRevD.102.115003} we studied neutrino decoherence resulting from neutrino interactions with VBH, and here we instead quantitatively assess the influence of lightcone fluctuations 

% \todo{2001.06016 also discusses fluctuating step size}

% \todo{Asymmetric fluctuations}

\subsubsection{Simulating neutrinos propagating in fluctuating space-time}
\label{sec:decoh_mc}

To test the influence of space-time fluctuations on neutrino propagation and the resulting neutrino decoherence, we implement a simulation of propagating neutrino states and stochastically inject travel distance fluctuations. This simulation software is also described in our previous study of neutrino decoherence from $\nu$-VBH interactions~\cite{PhysRevD.102.115003}. The neutrino mass states are propagated in discrete distance steps, with the states given by:

\begin{equation}
\label{eqn:plane_wave_perturbed}
\ket{\nu_{j}(L)} = \exp{ -i \left( \frac{ m_j^2 }{ 2 E } \left[ L + \Delta L(L) \right] \right) } \ket{\nu_{j}(0)} ,
\end{equation}

\noindent where $\ket{\nu_{j}}$ is the neutrino mass state $j$ ($j=1,2,3$ in the $3v$ paradigm) of mass $m_j$, with $E$ being the neutrino energy. Our lightcone model (defined in Section \ref{sec:lightcone_fluctuations}) specifies the uncertainty, $\deltaLref$, of each distance $\Lref$ travelled by a particle, which we represent in these simulations by evolving the neutrino states in discrete distance steps of size $L^{\prime}_0$, where the value of $L^{\prime}_0$ is a random number drawn from a normal distribution with mean $\Lref$ and standard deviation $\deltaLref$. The accumulated distance travelled is the sum of these steps, given by $L^\prime = \sum L^\prime_0 = L + \Delta L(L)$, where $L$ is the travel distance in the absence of fluctuations and $\Delta L(L)$ is the accumulated change in distance for a particular neutrino. This expression yields standard neutrino propagation when $\deltaLref = 0$ (and thus $\Delta L(L) = 0$).

The flavor transition probability after a given distance is determined by rotating the current state to the neutrino flavor basis, as defined by the Pontecorvo-Maki-Nakagawa-Sakata (PMNS) mixing matrix~\cite{Pontecorvo:1957qd, Maki:1962mu}, $U$, and projecting onto the desired final flavor state according to: 

\begin{equation}
\label{eqn:osc_prob_projection}
P( \nu_\alpha \rightarrow \nu_\beta ) \equiv P_{\alpha \beta} = |\braket{\nu_{\beta}(L)}{\nu_{\alpha}(0)}|^2 ,
\end{equation}

\noindent where $\alpha,\beta$ represent flavor indices ($e,\mu,\tau$ in the $3v$ paradigm). 

To probe the phenomenology of this system, we first test a 2-state system with toy (e.g. unrealistic) parameters chosen for clear visualisation (listed in \Cref{table:2nu_params}). Two fluctuation scenarios are considered. In the first case, each distance step is fluctuated independently, e.g. the fluctuations are uncorrelated as considered in the $m=1/2$ scenario described in Section \ref{sec:distance_dependence}. In the second case, the first step is fluctuated randomly as for the uncorrelated case, but all subsequent steps feature are fluctuated by the same amount. This represents a fully-correlated ($m=1$) distance fluctuation scenario (or equivalently a scenario where velocity instead fluctuates).

\begin{table}[htp]
 \begin{tabular}{|c| c|} 
 \hline
 Parameter & Value \\
 \hline
 \hline
 \# states & 2 \\ \hline
 Mixing angle, $\theta$ & 30\degree \\ \hline
 $\lambda$ & 200\,km \\ \hline
 $\Delta m^2$ & 0.012\,$\rm{eV}^2$ \\ \hline
 $E$ & 1\,GeV \\ \hline
 $\Lref$ & 1 km \\ \hline 
 Initial flavor & $\nu_\alpha$ \\ \hline 
 $\Lcoh$ & 500 km \\ \hline 
\end{tabular}
\caption{Parameters used for the propagating 2-state system. The mass states are labelled $0,1$ and the flavor states $\alpha,\beta$. The parameter values are chosen to produce clear demonstrations of the behaviour, rather than to represent realistic neutrino parameters. The mass splitting $\Delta m^2$ is chosen to give the desired oscillation wavelength $\lambda$.}
\label{table:2nu_params}
\end{table}

\begin{figure}[htp]
    \centering
    \includegraphics[trim=0.cm 0.0cm 0.0cm 0.0cm, clip=true, width=\linewidth]{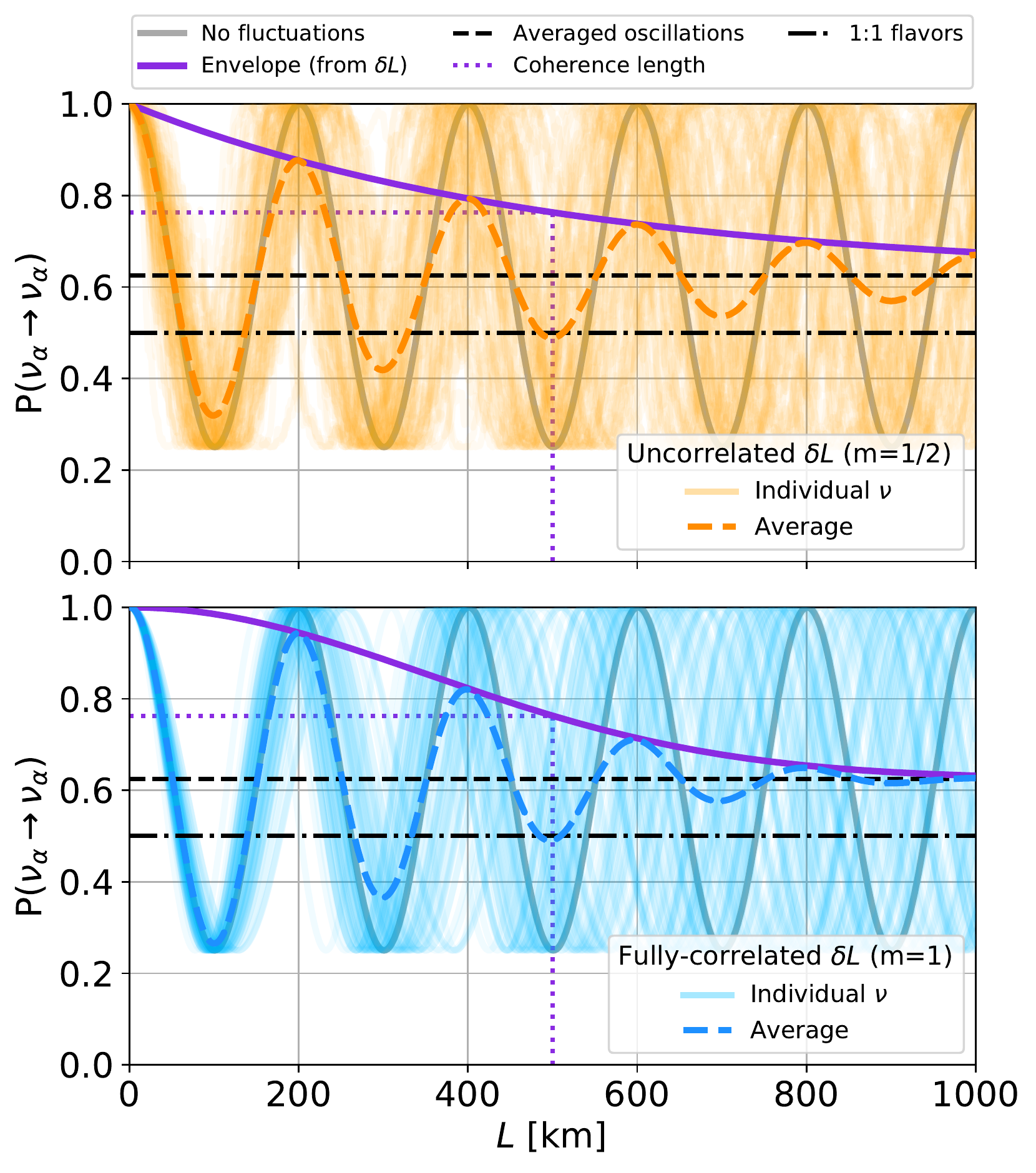}
    \caption{Decoherence in a MC simulation of neutrino propagation in the presence of lightcone fluctuations, resulting in the damping of neutrino oscillations. The upper panel shows an uncorrelated fluctuation scenario whilst the lower panel shows fully-correlated fluctuations. In both cases $\deltaLref$ is chosen such that the coherence length is the same for both. A 2-flavor system is shown with toy parameters selected for clarity, see \Cref{table:2nu_params}. }
    \label{fig:toy_model_2flav}
\end{figure}

The neutrino survival probabilities resulting from these simulations in both the the uncorrelated and fully-correlated scenarios are shown in upper and lower panels of \Cref{fig:toy_model_2flav} respectively. In both cases, the translucent coloured lines represent individual neutrinos, whilst the dashed colored line shows the average behaviour of the neutrino ensemble. It is this average behaviour which a neutrino counting experiment is ultimately sensitive to. $\deltaLref$ is chosen for the two cases such that they have the same \textit{coherence length} (see Section \ref{sec:decoh_analytic_2flav}), which in practice means a far smaller step size fluctuation for the fully-correlated case since the accumulation effect is much stronger. In both cases the expected damping of oscillations that is characteristic of neutrino decoherence is clearly observed, verifying that decoherence does indeed result from lightcone fluctuations.

We see in \Cref{fig:toy_model_2flav} that for both scenarios the large $L$ limit (e.g. when coherence is completely lost at large distances) is the average of the unfluctuated oscillations, given by $\overline{P_{\alpha \beta}} = \sum_j |U_{\alpha j}|^2 |U_{\beta j}|^2$. This is distinct from so-called \textit{relaxation} scenarios~\cite{GUZZO2016408, Gomes:2020muc} where the limiting case is equal populations of all flavors, as was identified in our previous work for certain $\nu$-VBH interaction models~\cite{PhysRevD.102.115003}. Differences in these large $L$ limits can in principal be used to distinguish between decoherence scenarios should a signal be detected.

An important distinction between the uncorrelated and fully-correlated cases is the functional form of the damping, visualised by the purple damping envelopes shown in \Cref{fig:toy_model_2flav}. This is expected given the differing distance dependence of the two scenarios. For the uncorrelated case, the envelope follows a $e^{-L}$ trend, whilst for the fully-correlated case we instead see damping of the form $e^{-L^2}$. This is explored further in the next section.

% TODO I think \href{https://users.aber.ac.uk/ruw/teach/260/ft3.php}{this page about Fourier transforms probably has the answer about why my sums of wave ultimately damp. But still need to make sense of it.}

\subsubsection{Connecting the simulations and distance fluctuation parameterisation}
\label{sec:decoh_analytic_2flav}

We now seek an analytic description of the decoherence phenomenon observed in the simulations presented in Section \ref{sec:decoh_mc}, and by extension neutrino decoherence from lightcone fluctuations more generally. This description should relate the damping effects to the underlying distance fluctuations parameterised by \Cref{eq:deltaL}.

The damping effect occurs as the spread in neutrino travel distances due to lightcone fluctuations grows, and the effect is expected to become large when $\deltaL \sim \lambda$, where $\lambda$ is the oscillation wavelength. Given that we observe $e^{-L}$ damping for the uncorrelated $\deltaL \propto L^{1/2}$ (e.g. $m=1/2$) case and $e^{-L^2}$ damping for the fully-correlated $\deltaL \propto L$ (e.g. $m=1$) case, this implies a damping envelope of the form:

\begin{equation}
\exp{ - \left( \frac{\delta L}{\lambda} \right)^2 } ,
\label{eq:envelope_vs_deltaL_propto}
\end{equation}

which given \Cref{eq:deltaL} yields:

\begin{equation}
\exp{ - \left[ \frac{ \deltaLref }{\lambda} \left( \frac {L} {\Lref} \right)^m \left( \frac {E} {\Eref} \right)^n \right]^2 } .
\label{eq:envelope_vs_L_propto}
\end{equation}

Damping envelopes of this form are shown by purple solid curves in \Cref{fig:toy_model_2flav}, defined as:

\begin{equation}
P_{\alpha \alpha} = \overline{P_{\alpha \alpha}} + \left( 1 - \overline{P_{\alpha \alpha}} \right) \exp{ - \left( \frac{\delta L}{ \eta \lambda} \right)^2 } ,
\label{eq:envelope_vs_deltaL}
\end{equation}

\noindent where $\eta$ is a $\order{1}$ dimensionless constant of proportionality defined such that the damping term is $e^{-1}$ when $\deltaL = \eta \lambda$, and can be thought of as defining the fraction of the oscillation wavelength that the distance uncertainty must accumulate to in order to produce strong decoherence effects. The value of $\eta$ will depend on the specific functional form of the fluctuations, and for the normally distributed step size fluctuations in our simulations we find $\eta \sim 0.23$ (e.g. when $\deltaL$ is comparable to a quarter of the wavelength). In common with other works~\cite{PhysRevD.102.115003}, we define the distance after which the damping term is $e^{-1}$ as the \textit{coherence length}, $\Lcoh$, which is given by:

\begin{equation}
\Lcoh = \Lref \left(\frac{\eta \lambda}{\deltaLref}\right)^{\frac{1}{m}} \left(\frac{\Eref}{E}\right)^{\frac{n}{m}} .
\end{equation}

The $\deltaL^2$ dependence observed in \Cref{eq:envelope_vs_deltaL_propto,eq:envelope_vs_L_propto,eq:envelope_vs_deltaL} can be understood by noticing that distance fluctuations are equivalent to frequency fluctuations for a sine wave, e.g. $
\Delta \omega \equiv ( \Delta L / L ) \omega \implies \omega \left[ L + \Delta L (L) \right] = \left[ \omega + \Delta \omega (L) \right] L$, where $\omega$ is the angular frequency of the wave. The sum of an infinite series of sine waves with differing frequencies but common amplitude and phases indeed features the same squared damping effect, and is directly analogous to the case of a neutrino propagating in fluctuating space-time. 

\subsubsection{Analytic decoherence operator}
\label{sec:decoh_analytic_3flav}

Now that we have expressed the damping effects we observe in these simulations in terms of our distance fluctuation parameterisation, we proceed to define a full decoherence operator suitable for describing neutrino propagation in fluctuating space-time. 

Neutrino decoherence is often represented using an open quantum system formalism~\cite{Benatti_2000, gago2002study, PhysRevLett.85.1166, PhysRevD.96.093009, Mavromatos:2006yy, Buoninfante:2020iyr, PhysRevD.99.075022, OHLSSON2001159, Farzan:2008zv, Coloma:2018idr, Carpio:2018gum, Carpio:2017nui, Anchordoqui:2005gj, PhysRevD.95.113005, PhysRevD.91.053002, PhysRevD.76.033006, PhysRevLett.118.221801, GUZZO2016408, Morgan:2004vv, Abbasi:2009nfa, Nieves:2019izk, Gomes:2020muc, Ohlsson:2020gxx, PhysRevD.101.056004, de_Holanda_2020, Nieves:2020jjg} considering both the neutrino and its environment, and beyond neutrinos this formalism has also be employed to study decoherence resulting from gravitational sources more generally~\cite{Anastopoulos:2013zya, Oniga:2017pyq, Bassi:2017szd}. The stochastic processes we consider in this work cause our knowledge of the neutrino to degrade over time, which in the language of open quantum systems constitutes the evolution from an initially \textit{pure} quantum state to a \textit{mixed} quantum state. Both mixed and pure quantum states can be mathematically expressed using the density matrix formalism, where the density matrix, $\rho$, for a system of $j$ states $\psi_j$ of probability $p_j$ is given by: 

\begin{equation}
\label{eqn:density_matrix}
\rho = \sum_j p_j \ket{\psi_j} \bra{\psi_j} .
\end{equation}

The time (or equivalently distance) evolution of an open quantum system is given by the Lindblad master equation~\cite{lindblad1976}:

\begin{equation}
\label{eqn:decoh_master}
\dot{\rho} = -i[H,\rho] - \Drho ,
\end{equation}

\noindent where $H$ is the Hamiltonian of the system (in which conventional oscillation effects are encoded) and $\Drho$ is a \textit{decoherence operator} defining stochastic/decoherence effects in the system. For a 3-flavour neutrino system (e.g. Nature as we currently know it), $\rho$, $H$ and $\Drho$ are $3 \times 3$ matrices. The neutrino flavor transition probability for such a system is given by:

\begin{equation}
\label{eqn:density_matrix_transition_prob}
P_{\alpha \beta} = \mathrm{Tr}[\rho_\alpha(t)\rho_\beta(0)] .
\end{equation}

The operator $\Drho$ encodes the decoherence effects in the system. In many existing studies simple forms for this operator have been assumed with manageable numbers of free parameters to test against experimental data, although in some cases the decoherence effects have been derived from first principals (see e.g.~\cite{Nieves:2019izk, PhysRevD.101.056004, Nieves:2020jjg}). In this week we seek to determine the form of $\Drho$ representing the distance fluctuations we are considering, and ultimately produce the damping effects we observe. From \Cref{eq:envelope_vs_L_propto} we see the need for a solution to \Cref{eqn:decoh_master} of the general form $\rho  \propto \exp{ - L^{2m} }$ implying a form $\Drho \propto - 2m L^{2m-1} \rho$, which once differentiated yields the desired damping form. Taking into account the full distance fluctuation parameterisation \Cref{eq:deltaL}, for a 3-flavour system the decoherence operator is:

\begin{widetext}
\begin{equation}
\label{eqn:Drho_lightcone_fluctuations}
\renewcommand{\arraystretch}{2}
\Drho =  \frac{ 2m (\delta L_0)^2 L^{2m-1} }{ L^{2m}_0 } \left( \frac{E}{\Eref} \right)^{2n} \begin{pmatrix}
0 & \dfrac{\rho_{21}}{ (\eta \lambda_{21})^2}  & \dfrac{\rho_{31}}{ (\eta \lambda_{31})^2} \\ 
\dfrac{\rho_{21}}{ (\eta \lambda_{21})^2} & 0 & \dfrac{\rho_{32}}{ (\eta \lambda_{32})^2} \\ 
\dfrac{\rho_{31}}{ (\eta \lambda_{31})^2} & \dfrac{\rho_{32}}{ (\eta \lambda_{32})^2} & 0 \\ 
\end{pmatrix} ,
\end{equation}
\end{widetext}

\noindent where $\lambda_{ij}$ is the oscillation wavelength corresponding to the mass splitting $\Delta m^2_{ij}$, where $\lambda_{ij} = 4 \pi E / \Delta m^2_{ij}$ in vacuum. There are three wavelengths to consider here, rather than a single wavelength for the 2-flavour system considered in Section \ref{sec:decoh_analytic_2flav}.

\Cref{eqn:Drho_lightcone_fluctuations} is one of the primary results of this work, and provides an operator characterising the general case of neutrino decoherence from lightcone fluctuations, including those resulting from fluctuating space-time models. This allows neutrino transition probabilities to be computed given some underlying distance fluctuation parameters, e.g. $\{\Lref, \deltaLref, m, n\}$, which can then be tested against experimental data. Alternatively, existing constraints on neutrino decoherence can be re-interpreted in terms of this underlying model, and the results compared to corresponding constraints on space-time fluctuations from $\gamma$-ray observations~\cite{Vasileiou2015, Perlman_2015, Cooke:2020rco}.

\begin{figure}[htp]
    \centering
    \includegraphics[trim=0.cm 0.0cm 0.0cm 0.0cm, clip=true, width=\linewidth]{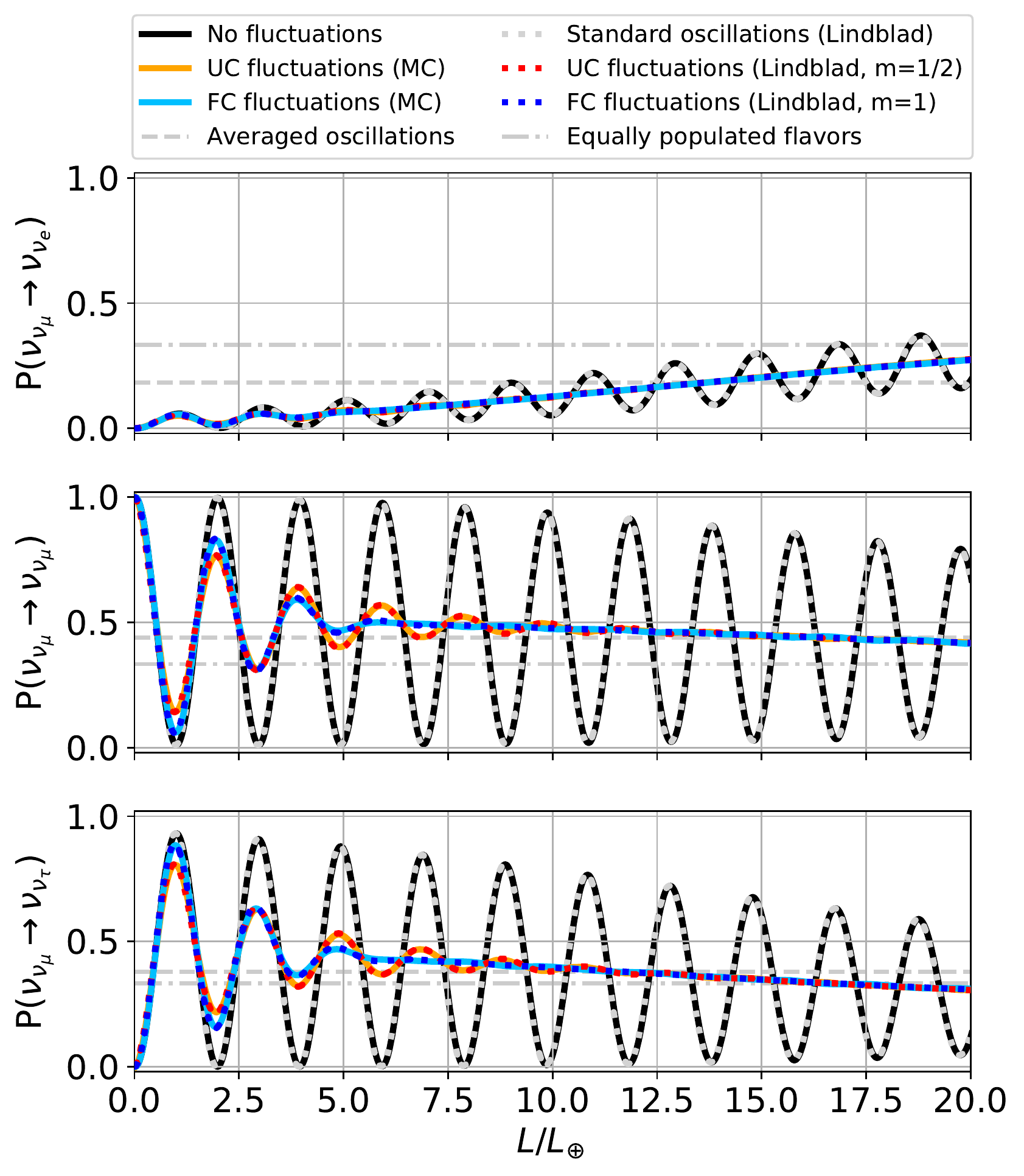}
    \caption{Decoherence in a MC simulation of neutrino propagation in the presence of lightcone fluctuations for a 3-flavour system with realistic mixing parameters in the atmospheric neutrino parameter space (see \Cref{table:3nu_params}). Both uncorrelated (UC) and fully-correlated (FC) fluctuations are shown. Solid lines show simulation results, whilst dotted lines show the corresponding analytic expression computed using $\Drho$. The travel distance is expressed as the number of Earth diameters, $L_\oplus \sim$ 12,700 km, traversed (an atmospheric neutrino experiment is only sensitive to neutrinos crossing a single diameter). Matter effects are not included.}
    \label{fig:toy_model_vs_lindblad_3flav}
\end{figure}

To verify and demonstrate this operator, we now simulate distance fluctuations as described in Section \ref{sec:decoh_analytic_2flav} but for a full 3-flavour system with realistic neutrino mixing parameters (listed in \Cref{table:3nu_params}), and compare the resulting neutrino transition probabilities with those computed using $\Drho$. \Cref{fig:toy_model_vs_lindblad_3flav} shows these results for the $\nu_\mu \rightarrow \nu_{e,\mu,\tau}$ channels, where the neutrino energy and baseline are chosen to be representative of atmospheric neutrinos. Both fully-correlated and uncorrelated fluctuations are shown, with the injected step size fluctuation (e.g. $\deltaLref$) chosen such that both share a common coherence length with respect to $\lambda_{31}$. The Lindblad analytic form exactly matches the simulation results, again with $\eta \sim 0.23$ as was the case in the 2-flavour system.

% \noindent \textbf{Derivation:}

% Want to find a $\Drho$ that produces something like this (e.g. the damped state):

% \begin{equation}
% \rho  \propto \exp{ - L^{2m} }
% \end{equation}

% $\Drho$ appears in the master equation like this (neglecting standard oscillations momentarily):

% \begin{equation}
% \frac{d\rho}{dL}  = - \Drho
% \end{equation}

% So basically I want to define a $\Drho$ that integrates w.r.t. $L$ to give me my desired result. This derivation is thus:

% \begin{gather*}
% d\rho= - \Drho \, dL \\
% \Drho = - 2m L^{2m-1} \rho \\
% \implies d\rho = - 2m L^{2m-1} \rho \, dL \\
% \frac{d\rho}{\rho} = - 2m L^{2m-1} \, dL \\
% \ln{\rho} = - \frac{ 2m L^{2m} }{ 2m } \\
% \rho = \exp{ - L^{2m} }
% \end{gather*}

\begin{table}[htp]
 \begin{tabular}{|c| c|} 
 \hline
 Parameter & Value \\
 \hline
 \hline
 $\Delta m^2_{21}$ & $7.39 \times 10^{-5}$ $\rm{eV}^2$ \\ \hline
 $\Delta m^2_{31}$ & $2.528 \times 10^{-3}$ $\rm{eV}^2$ \\ \hline
 Mass ordering & Normal \\ \hline
 $\theta_{12}$ & 33.82$\degree$ \\ \hline
 $\theta_{13}$ & 8.60$\degree$ \\ \hline
 $\theta_{23}$ & 48.6$\degree$ \\ \hline
 $\delta_{CP}$ & 221$\degree$ \\ \hline
 $E$ & 25 GeV \\ \hline
 $L_{\rm{coh},31}$ & $3 L_\oplus$ \\ \hline 
\end{tabular}
\caption{Parameters used for evaluating atmospheric neutrino oscillations. Neutrino oscillation parameters are taken from NuFit 4.1 global fit results (normal mass ordering, SuperKamiokande data included)~\cite{NuFit41}.}
\label{table:3nu_params}
\end{table}

One interesting aspect of decoherence resulting from lightcone fluctuations is the differing coherence lengths for different oscillation frequencies. In \Cref{fig:toy_model_vs_lindblad_3flav} the damping of the higher frequency oscillations resulting from the atmospheric mass splitting $\Delta m^2_{31/2}$ (with $\lambda \sim 2 L_\oplus$) is clearly seen, with an injected coherence length of $\Lcoh = 3 L_\oplus$. However, the flavour transition probability still continues to change with distance due to the lower frequency oscillations resulting from the solar mass splitting, $\Delta m^2_{21}$. 

\begin{figure}[htp]
    \centering
    \includegraphics[trim=0.cm 0.0cm 0.0cm 0.0cm, clip=true, width=\linewidth]{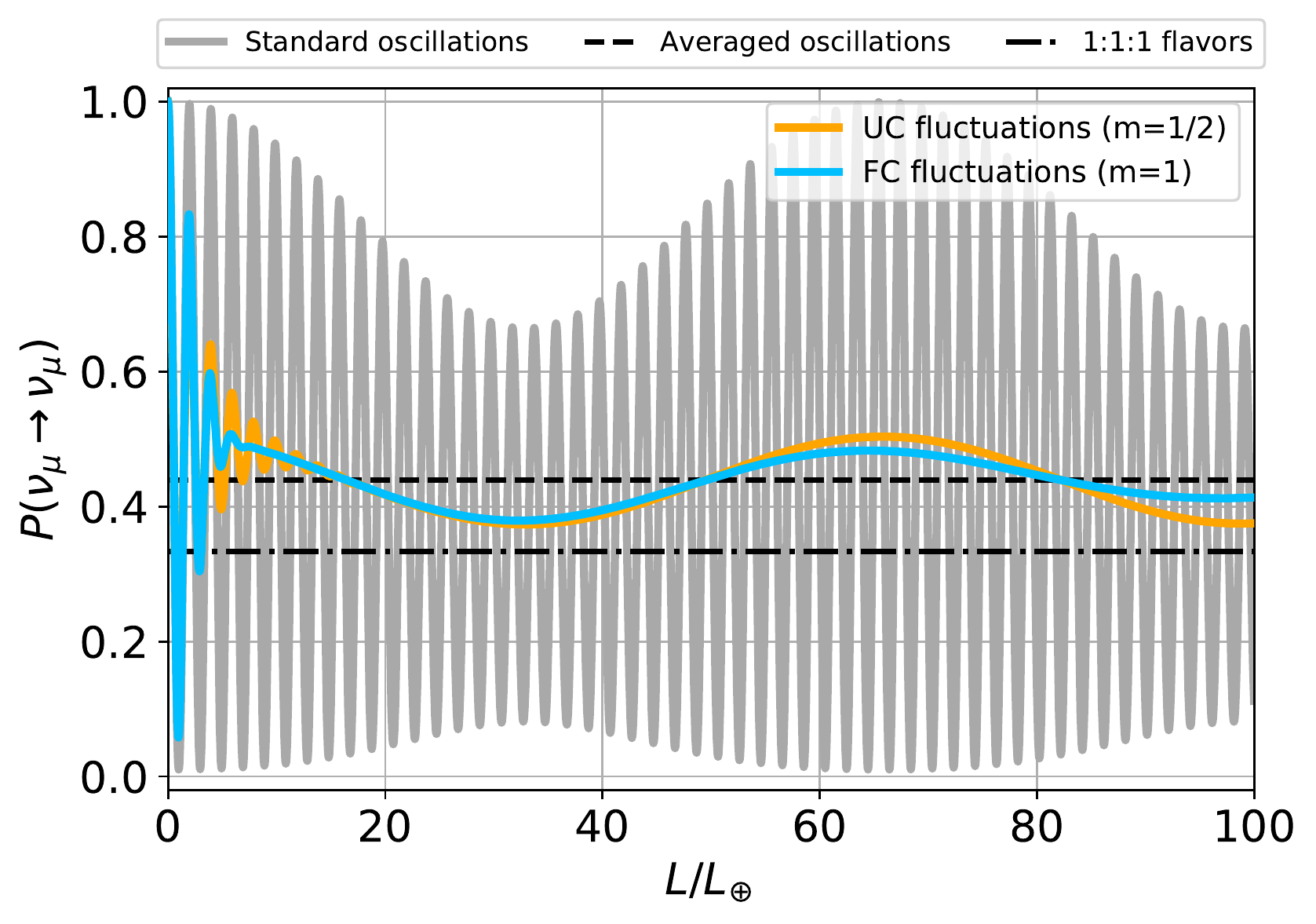}
    \caption{The $\nu_\mu$ survival probability as shown in the central panel of \Cref{fig:toy_model_vs_lindblad_3flav}, but shown over a longer distance such that the lower frequency oscillations of resulting from $\Delta m^2_{21}$ can be seen. Only the analytic Lindblad curves are shown.}
    \label{fig:lindblad_3flav_long}
\end{figure}

\Cref{fig:lindblad_3flav_long} shows the $\nu_\mu$ survival channel over larger distances where this second (solar) oscillation frequency is clearly visible even after the first (atmospheric) frequency has damped. Even over the larger distance these lower frequency oscillations have not damped, although it can be seen that the fully-correlated case is damping more quickly with distance than the uncorrelated case (despite having identical coherence lengths for the higher frequency oscillations), which is expected due to the differing distance dependence of the damping terms ($e^{-L^2}$ and $e^{-L}$ respectively). The large difference in coherence length between the two oscillation frequencies is a consequence of the orders of magnitude difference between the solar and atmospheric mass splittings.

This difference between the coherence length of different oscillation frequencies differs from the $\nu$-VBH interaction case we considered in our previous work, which produced uniform damping in all channels. This therefore provides a potential discriminating factor between different scenarios should a decoherence signal be discovered experimentally.

Another key difference between decoherence from lightcone fluctuations and other possible sources relates to the energy dependence. The $\lambda_{ij}^{-2}$ dependence of the damping effects results in an intrinsic $E^{-2}$ dependence (since $\lambda \propto E$). This means that in the absence of any explicit energy dependence (e.g. $n \neq 0$) in the system -- for example the suppression of $\deltaL$ below the Planck scale ($n > 0$) -- lower energy neutrinos offer greater sensitivity to these decoherence effects. Indeed, only in cases with $n > 1$ (given the $E^{2n}$ term) do the decoherence effects start to grow with neutrino energy.  

\subsubsection{Sensitivity to natural Planck scale effects}

We have established that neutrino decoherence results from lightcone fluctuations, and shown that the resulting damping effects become large when $\deltaL(L) \sim \lambda_{ij}$. We now consider the energies and baselines of neutrinos observed from various sources to determine where a potential signal would be expected to be strongest. Given that evidence of fluctuating space-time has not yet been observed, such effects likely only manifest over very large distances, and/or are suppressed at energies below the Planck scale.

\begin{figure}[htp]
    \centering
    \includegraphics[trim=0.0cm 0.0cm 0.0cm 0.0cm, clip=true, width=\linewidth]{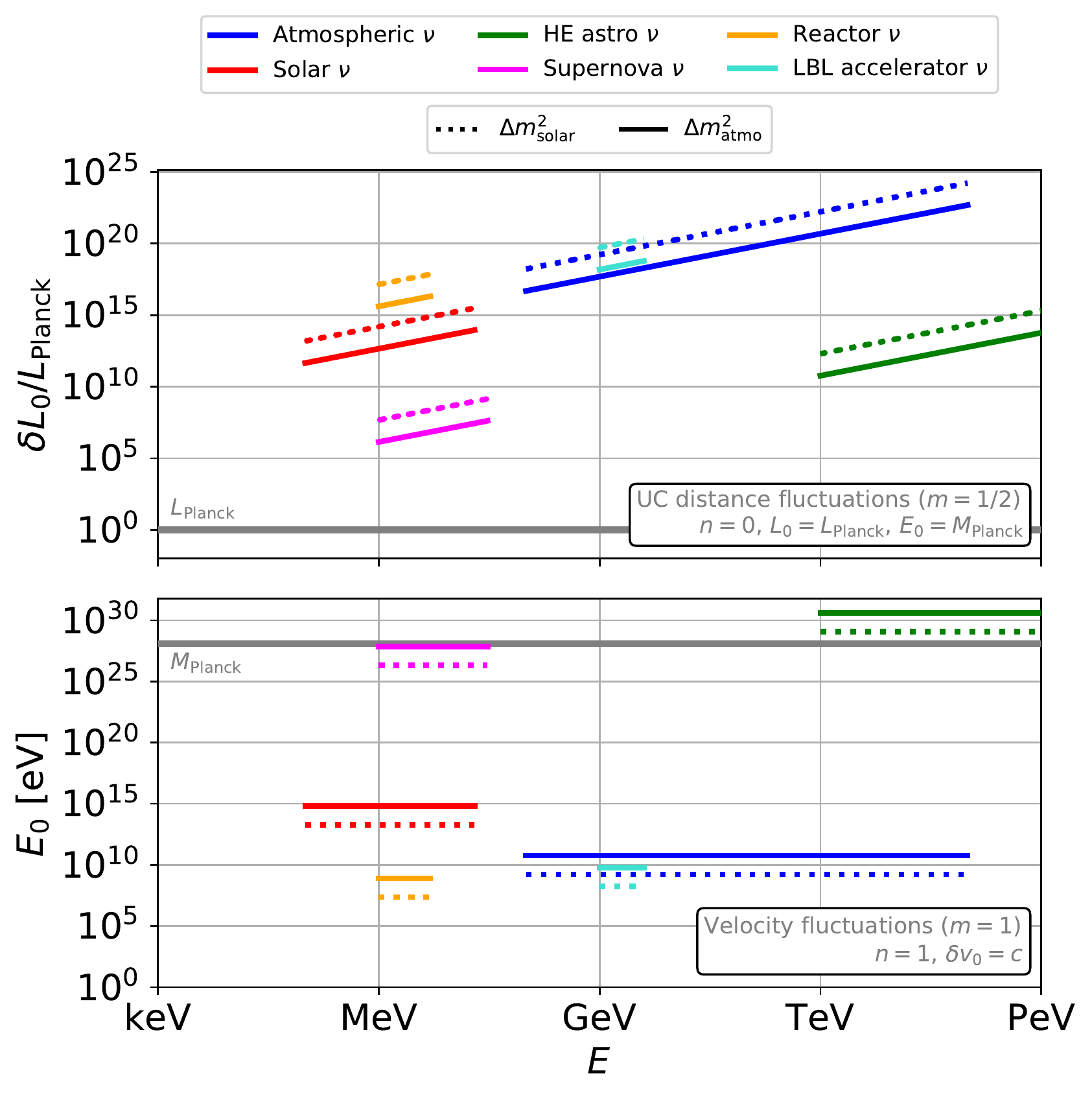}
    \caption{Underlying model parameters producing strong decoherence effects for neutrinos from a range of sources. The upper panel shows the travel distance fluctuation requirement for each Planck length traversed in the case of uncorrelated, energy independent distance fluctuations. The lower panel shows the new physics energy scale required to produce strong decoherence effects in a velocity fluctuation scenario with energy suppression $\propto E/\Mplanck$. Upper limits for travel distance are assumed, e.g. one Earth diameter for atmospheric neutrinos, etc. }
    \label{fig:decoherence_sensitivity}
\end{figure}

\Cref{fig:decoherence_sensitivity} shows the underlying model parameters required to produce strong decoherence effects for a variety of neutrino sources. Damping of both the higher (atmospheric) and lower (solar) frequencies are shown. The upper panel considers the case of energy independent uncorrelated distance fluctuations, for which sensitivity to natural Planck scale effects (e.g. $\deltaLref = \Lplanck$) has been achieved using astrophysical photon observations~\cite{Perlman_2015, Cooke:2020rco}. The y axis indicates the size of fluctuations required for each Planck length travelled to produce strong decoherence effects, which is inversely proportional to energy in this case due to the wavelength-dependence. We see that for all neutrinos sources tested, even those with cosmological baselines, the required fluctuations is orders of magnitude larger than the natural expectation, giving little prospect of a signal detection unless lightcone fluctuations significantly exceed this natural expectation. This is a consequence of the macroscopic oscillation wavelengths, whereas the microscopic wavelengths of high energy photons makes them more susceptible to the effects of lightcone fluctuations. The prospects for a neutrino signal would be further reduced in the case of any energy-suppression ($n>0$) of the effects.

The lower panel of \Cref{fig:decoherence_sensitivity} instead shows the case of distance independent velocity fluctuations as discussed in Section \ref{sec:velocity_fluctuations}, suppressed by a single energy power ($n=1$). Natural Planck scale limits for such a scenario have been achieved by constraining the arrival time spread of high energy photons from a short GRB~\cite{Vasileiou2015} (see Section \ref{sec:propagation_time_fluctuations}). In this case we vary the energy scale $\Eref$ of the new physics producing the velocity fluctuations, where the natural expectation for quantum gravity is $\Eref = \Mplanck$. We see that significant decoherence effects are indeed expected in this scenario for neutrinos travelling cosmological and possibly even galactic baselines, yielding a possible detection channel.

However, there are major challenges in observing such a signal. The majority of high energy astrophysical neutrinos, observed by the IceCube neutrino observatory~\cite{Aartsen:2013jdh}, have not been associated with a particular source, but instead appear as an approximately isotropic diffuse flux. This flux likely results from many individual sources at unknown distances, and is thus incoherent even in the absence of lightcone fluctuations, producing an oscillation averaged flavor composition at the Earth exactly as would be expected in the lightcone fluctuation signal case\footnote{This is contrary to decoherence scenarios resulting in equal flavour populations, which can in principal be distinguished even with an incoherent source~\cite{PhysRevD.102.115003}.}. This is further compounded by the finite energy resolution of IceCube and other neutrino telescopes, which also degrades and averages the oscillation signal~\cite{OHLSSON2001159}. The only hope for detecting such a signal would therefore be the observation of neutrinos from a coherent astrophysical source (e.g. with a compact emission region compared to the oscillation wavelength).

We therefore see that the prospects of detecting neutrino decoherence from natural Planck scale lightcone fluctuations via flavour-based measurements do not look promising, given the macroscopic scale of oscillation wavelengths and challenges in observing decoherence in astrophysical neutrinos. We now also consider an alternative potential detection channel.

%----------------------------------------------------------------
%----------------------------------------------------------------
%----------------------------------------------------------------

\subsubsection{Comparison to other studies}
\label{sec:comparison_to_other_work}

In this work we have considered scenarios where the consequence of fluctuating/uncertain space-time is fluctuations in the travel distance/time between two points, and the resulting decoherence effects in propagating neutrinos. In \cite{Mavromatos:2006yy}, an alternative but comparable model is studied where the space-time metric itself experiences fluctuations, also resulting in lightcone fluctuations and neutrino decoherence. An analytic treatment is applied to quantify the average damping effects, as opposed to the simulation-based methods employed here.

The study considers fluctuations of a $(1+1)$D metric tensor (one time dimension and one spatial dimension, aligned with the particle travel direction) of the form:

\begin{equation}
g^{\prime} = O g O^T ,
\end{equation}

\begin{equation}
O = \begin{pmatrix}
a_1 + 1 & a_2 \\
a_3 & a_4 + 1
\end{pmatrix} ,
\end{equation}

where $a_i$ represent perturbations to the metric, which are Gaussian random variables with an average value of zero and standard deviation $\sigma_i$. $\sigma_i$ are free parameters characterising the fluctuations and can be considered the analogue of $\deltaLref$ in our work. $g$ and $g^{\prime}$ are the unfluctuated and fluctuated metric tensors respectively, with $g$ being taken as the Minkowski metric representing flat space-time.

The case $\sigma_1 = \sigma_2 = \sigma_3 = 0$, $\sigma_4 > 0$ corresponds to pure distance fluctuations along the particle direction of travel, which is directly comparable to the model we have proposed in this work. The resulting damping term (in the two flavour neutrino system considered) has the form\footnote{\cite{Mavromatos:2006yy} also considers possible matter effects resulting from neutrinos interacting with a VBH background, characterised by a potential, $V$. This is a distinct effect from the model considered in this article, and we neglect these terms in this comparison (e.g. set $V=0$). These effects can however be compared to our previous study on neutrino-VBH interactions~\cite{PhysRevD.102.115003}.}:

\begin{equation}
\exp{ - \frac{ \left( \Delta m^2 \right)^2 }{ 2 E^2 } \sigma_4 L^2 } \sim \exp{ - \frac{ 1 }{ \lambda^2 } \sigma_4 L^2 } .
\label{eq:metric_fluc_damping}
\end{equation}

This is very similar to the damping term in the two flavour scenario derived in this work in the case where $m=1$ (and $n=0$, since \cite{Mavromatos:2006yy} does not consider energy-dependence in the fluctuations), where \Cref{eq:envelope_vs_L_propto} becomes:

\begin{equation}
\exp{ - \frac{1}{\lambda^2} \left( \frac {\deltaLref} {\Lref} \right)^2 L^2  } .
\label{eq:envelope_vs_L_propto_energy_independent_m1}
\end{equation}

We thus seem that both approaches (distance vs. metric fluctuations, simulation vs. analytic methods) produce qualitatively the same neutrino decoherence effects, with $\exp{-L^2}$ damping\footnote{Referred to as \textit{Gaussian damping} in \cite{Mavromatos:2006yy}.} dependent on $1/\lambda^2$ and a parameter characterising the uncertain/fluctuating space-time. This agreement serves to verify both approaches.

\subsubsection{Comparison to wave packet decoherence}
\label{sec:comparison_to_wp_decoh}

Neutrino mass states propagate as wave packets, which physically separate over large enough distances due to their differing masses, degrading the superposition producing neutrino oscillations and resulting in the damping of flavour transitions and neutrino decoherence~\cite{Nussinov:1976uw}. The coherence length of neutrinos in current neutrino oscillation experiments is expected to far exceed the measurement baselines however, meaning such effects can typically be neglected.

It is however interesting to compare the decoherence resulting from lightcone fluctuations considered in this work to the case of wave packet decoherence, where the damping can be expressed as~\cite{Giunti:2003ax}:

\begin{equation}
\exp{ - \frac{ \left( \Delta m^2 \right)^2 }{ 32 E^4 } \frac{1}{\sigma_x^2} L^2 } \sim \exp{ - \frac{ 1 }{ \lambda^2 } \frac{1}{ E^2 \sigma_x^2} L^2 } ,
\label{eq:wp_decoh_damping_term}
\end{equation}

where $\sigma_x$ is the spatial width of neutrino wave packet along the direction of travel. Wave packet decoherence in curved space-time can also be considered, which for the case of a neutrino propagating along radial geodesics in the Schwarzschild metric yields nearly identical effects to the flat space-time case except that the travel distance is increased due to the space-time curvature, enhancing the damping effects~\cite{Chatelain:2019nkf}.

Comparisons between Equations (\ref{eq:wp_decoh_damping_term}) and (\ref{eq:envelope_vs_L_propto_energy_independent_m1}) indicate that wave packet decoherence is phenomenologically similar to the lightcone fluctuation decoherence considered in this work in the case of $m=1$ (e.g. fully-correlated fluctuations), with the damping term depending on both $L^2$ and $1/\lambda^2$. Notably however, wave packet decoherence features an intrinsic $1/E^2$ dependence (in addition to the energy-dependence from the $1/\lambda^2$ term) not present in the lightcone fluctuation scenario. The wave packet and lightcone fluctuation cases therefore become largely degenerate when the size of the lightcone fluctuations are assumed to have an extrinsic $n=-2$ energy-dependence, and the two scenarios cannot easily be separated. However, such an inverse energy-dependence is not typical of Planck scale physics scenarios, where effects are instead expected to be suppressed at lower energies ($n>0$). 

We therefore see that despite some phenomenological similarities, wave packet decoherence and lightcone fluctuation decoherence from Planck scale physics can in principal be distinguished via their energy- and distance-dependence, aside from the special case where $m=1, n=-2$ for the lightcone fluctuations.

\subsection{Propagation time fluctuations}
\label{sec:propagation_time_fluctuations}

Fluctuations in neutrino travel distance would correspondingly produce fluctuations in a particle's propagation time between two points, given by:

\begin{equation}
\delta t = \frac{\deltaL}{v} .
\label{eq:delta_t}
\end{equation}

Variations in the arrival time of cosmic messenger particles (mainly high energy photons) from cosmological sources have been searched for extensively in an effort to detect possible quantum gravity signals~\cite{Abdo2009, Vasileiou2015, ELLIS2019352, PhysRevD.102.063027, PhysRevD.99.083009, AMELINOCAMELIA2016318, Wei:2018ajw, HESS:2011aa}, with no signal observed to date. Searches have mainly focused on deterministic variations in particle velocity via modified dispersion relations, typically motivated by the prospect that Lorentz invariance symmetry or the weak equivalence principal is violated in quantum gravity. Neutrino-based constraints~\cite{ELLIS2019352, PhysRevD.102.063027, Wei:2018ajw} have more recently been derived from the multi-messenger observations of the first identified astrophysical high energy neutrino point source, the flaring blazar TXS 0506+056~\cite{IceCube:2018cha, IceCube:2018dnn}.

Stochastic modifications to particle propagation times as would result from lightcone fluctuations are less well explored. Planck scale constraints of velocity fluctuations have been derived~\cite{Vasileiou2015} from the arrival times of $\gamma$-rays from the distant GRB 090510~\cite{Ackermann_2010} in a weakly energy-suppressed scenario ($n=1$), but no limits of this kind exist for neutrinos.

Short GRBs are a powerful probe of particle travel time variations~\cite{AmelinoCamelia1998, AmelinoCamelia:2009pg}, since they have been detected at cosmological distances, display high energy photon emission in the keV-TeV energy range and crucially feature an initial prompt emission period of $\order{\rm{1\:s}}$. The duration of the observed emission in this prompt phase constrains variations in travel time due to lightcone fluctuations. The prompt emission additionally shows evidence of temporal sub-structure of $\order{\rm{\mu s - ms}}$~\cite{Ackermann_2010, Vasileiou2015} which can in principal yield even tighter constraints. 

GRBs are a candidate production site of high energy neutrinos~\cite{Aartsen_2016}, although to date there has been no significant correlation of GRBs with neutrinos\footnote{It has been show that GRBs can at most account for a small fraction of the diffuse astrophysical neutrino flux observed by the IceCube neutrino observatory~\cite{Aartsen_2016}. The source of the majority of this flux remains unknown. An alternative possibility is that lightcone fluctuations have prevented associations of $\gamma$-rays and neutrinos of GRBs in these studies due to arrival time fluctuations.}. Neutrino astronomy is still a new field however, and GRB neutrino emission, if detected, offers a number of advantages in searches for Planck scale physics compared to photons. Astrophysical neutrinos are observed at energies far in excess of that seen for photons, up to PeV~\cite{glashow_nature} and potentially EeV, and can therefore better overcome any suppression of quantum gravity effects below the Planck scale. Unlike neutrinos, high energy photons have a limited range due to absorption by the cosmic microwave background (CMB)~\cite{PhysRev.155.1408}. Additionally, the highest energy $\order{\rm{TeV}}$ photon emission observed from GRBs is typically not detected during the initial burst, but instead from subsequent longer duration processes (the so-called  \textit{afterglow})~\cite{MAGIC_GRB, HESS_GRB}. This is a consequence of the small field of view of the imaging atmospheric Cherenkov telescopes (IACT) used in these observations, meaning that they typically only observe GRBs when triggered by other telescopes with larger fields of view such as Fermi-LAT~\cite{Atwood_2009} (sensitive only to lower energy emission). The prompt emission is missed during the time taken to respond to the alert and point the telescopes, which is $\order{1 \: \rm{min}}$. This longer time scale emission is far less powerful in constraining arrival time fluctuations. Neutrino telescopes however typically observe the entire sky continuously, meaning that even prompt neutrino emission can be detected.

\begin{figure}[htp]
    \centering
    \includegraphics[trim=0.5cm 0.0cm 0.0cm 0.0cm, clip=true, width=\linewidth]{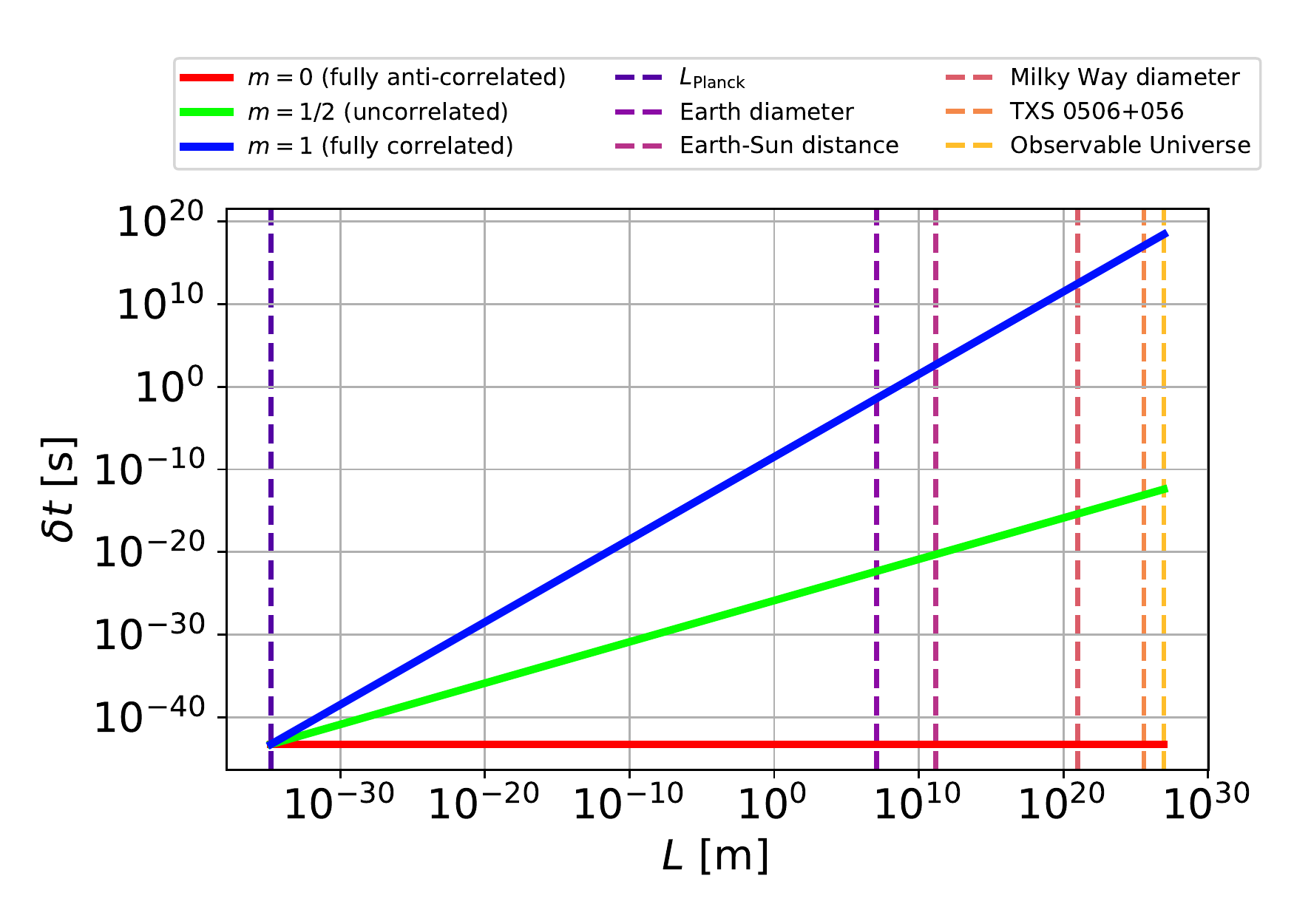}
    \caption{Propagation time variation expected for the `natural scenario' given by \Cref{eq:deltaL_natural} for a particle with $E=\Mplanck$ (or $n=0$), as a function of particle propagation distance. A number of reference distances are shown with dashed lines. Scenarios with differing $m$ are shown, with their interpretations discussed in Section \ref{sec:distance_dependence}.}
    \label{fig:delta_t_vs_L}
\end{figure}

To ascertain the potential of neutrino arrival time fluctuation searches to quantum gravity effects, \Cref{fig:delta_t_vs_L} shows the scale of time fluctuations vs. particle travel distance in the `natural' Plank scale distance fluctuation scenario discussed in Section \ref{sec:distance_uncertainty_natural}. This is the time analogue of the distance fluctuations shown in \Cref{fig:delta_L_vs_L}, and shows the case of a neutrino with Planck scale energy, or equivalently an energy independent ($n=0$) scenario. This is an upper limit on the size of the effects in energy-suppressed scenarios. 

For the uncorrelated distance fluctuation case ($m=1/2$), even over cosmological propagation distances the distance fluctuations only accumulate to $\leq \order{\rm{ps}}$ in scale, even before any possible energy-suppression is considered at observable particle energies. Since no source of high energy neutrino emission with such a short time scale is known, there are currently no real prospects of detecting such a scenario via arrival time fluctuations.

\begin{figure}[htp]
    \centering
    \includegraphics[trim=0.5cm 0.0cm 0.0cm 0.0cm, clip=true, width=\linewidth]{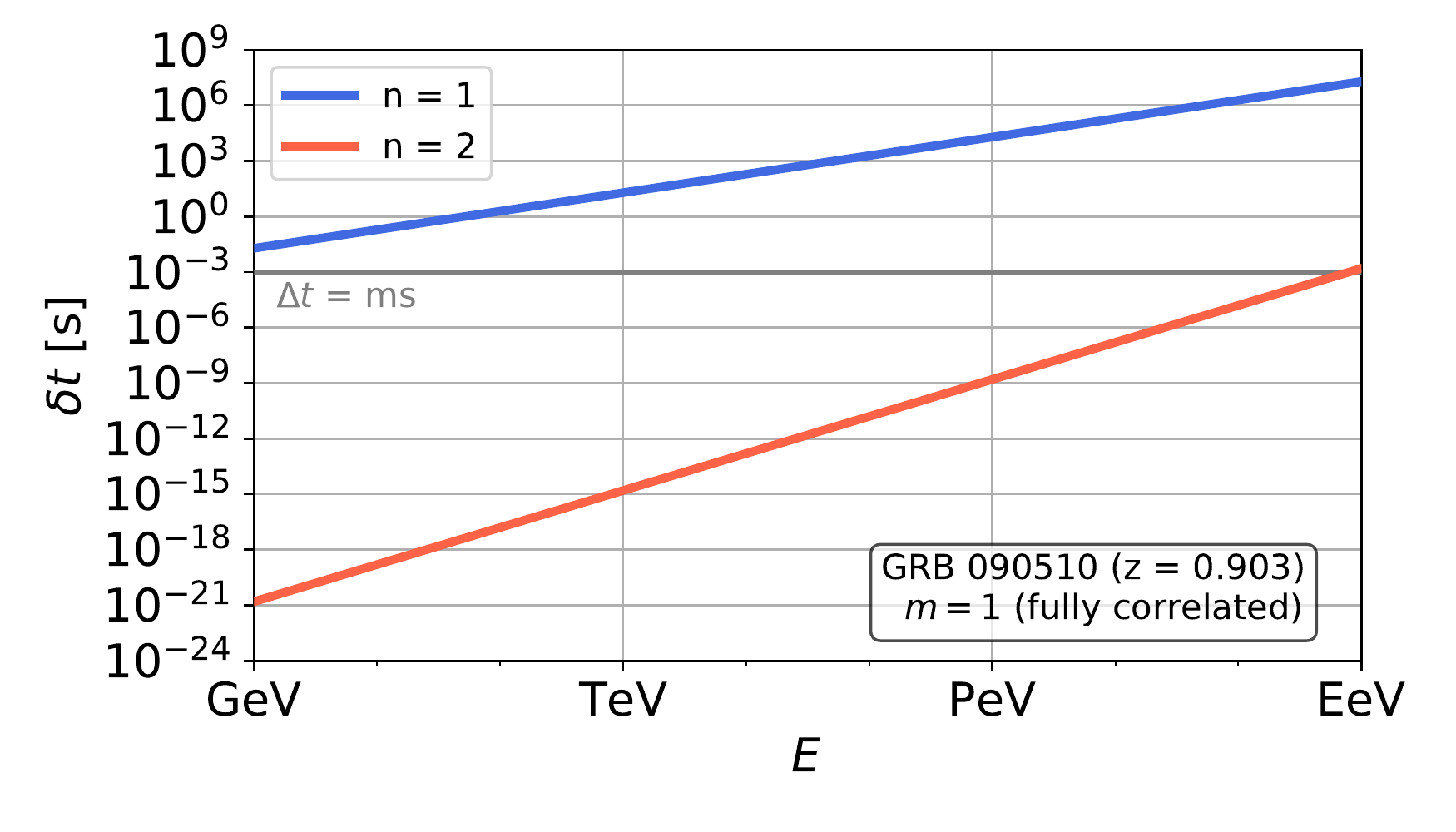}
    \caption{Propagation time variation expected for the `natural scenario' given by \Cref{eq:deltaL_natural} for a particle travelling from GRB 090510 to Earth, as a function of particle energy. Fully-correlated ($m=1$) distance fluctuations or equivalently distance independent velocity fluctuations are assumed. Both $E / \Mplanck$ (e.g. $n=1$) and $\left ( E / \Mplanck \right )^2$ (e.g. $n=2$) energy suppression scenarios are shown. The horizontal grey line indicates the approximate time sub-structure in the time profile of the prompt emission from the GRB, with $\delta t$ above this line yielding sensitivity to natural Planck scale physics.  }
    \label{fig:delta_t_vs_E_GRB}
\end{figure}

However, in velocity fluctuation ($m=1$) or correlated distance fluctuations ($m>1/2$) scenarios, the prospects are much improved. \Cref{fig:delta_t_vs_E_GRB} shows the arrival time fluctuations expected when $m=1$ as a function of particle energy for two energy-suppressed scenarios, $n=1, 2$. For particle travel distance the source is assumed to be GRB 090510 (redshift, $z = 0.903$~\cite{McBreen2010}), which has been the subject of a number of quantum gravity motivated arrival time studies~\cite{Abdo2009, Vasileiou2015} and has $\order{\rm{ms}}$ sub-structure in its prompt photon time distribution which would be degraded and ultimately unresolved if travel time fluctuations exceed this scale. 

From \Cref{fig:delta_t_vs_E_GRB} we see that travel time fluctuations would exceed this $\order{\rm{ms}}$ scale for $\gtrsim$GeV particles in the `natural' scenario shown when the energy suppression is $E / \Mplanck$ ($n=1$), which has enabled Planck scale limits on velocity fluctuations to be set using $\order{\rm{GeV}}$ $\gamma$-rays from this GRB~\cite{Vasileiou2015}. Should $\gtrsim$TeV neutrino emission be eventually detected from distant GRBs, we see that the sensitivity to Planck scale physics will significantly exceed that currently available from GeV photon observations, allowing the possible detection of quantum gravity effects beyond the reach of current measurements (for example if the energy scale of quantum gravity exceeds $\Mplanck$, or $\deltaLref < \deltaLplanck$). In fact, even longer duration sources (seconds or minutes) could still yield Planck scale physics signals for $>$TeV particles. Combined analyses of neutrino emission from multiple GRBs and multiple neutrino telescopes could be used to help overcome the low statistics inherent in neutrino point source observations when compared to $\gamma$-rays. 

For the $\left ( E / \Mplanck \right )^2$ ($n=2$) energy suppression scenario, the prospect of detecting Planck scale physics is far weaker, requiring observations of multiple EeV neutrinos from GRBs (for example by future large scale radio neutrino observatories). Since the Universe is opaque to photons at these energies however, neutrinos still offer the best detection prospect in this scenario.

% 2008.04323 Fig 1 shows E-L parameter space where photons are absorbed

Aside from the potential gains from the high energies observed in astrophysical neutrinos, constraining Planck scale physics with different messenger particles is also inherently desirable, as it is also possible that the effects of fluctuating space-time differ for different particle types~\cite{ELLIS2004669, PhysRevD.59.116008, Coleman:1998en}.

\section{Summary and conclusions}

In this work we have presented a heuristic parameterisation of particle propagation distance fluctuations, with the aim of probing the expectation that space-time fluctuates if gravity is a quantum force. This parameterisation accounts for both distance and energy dependence, unlike previous work that has considered only one or the other individually, and has also been shown to representative of velocity fluctuation scenarios.

The influence of these lightcone fluctuations on neutrino propagation was studied, considering both the loss of neutrino coherence (and corresponding damping of neutrino oscillations), and the broadening of neutrino arrival times from short duration distant astrophysical sources. Using simulations of propagating neutrino states in the presence of distance fluctuations, we have quantified the decoherence effects resulting from lightcone fluctuations, and determined an operator representing these effects in the framework of open quantum systems. This operator allows experimental searches for neutrino decoherence to be connected to potential underlying fluctuations in space-time, and compared to results from $\gamma$-ray astronomy.

Due to their macroscopic oscillation wavelengths, we have seen that neutrinos only experience significant decoherence effects in a small number of (optimistic) lightcone fluctuation scenarios, and even then only over cosmological or perhaps galactic distances. The incoherent nature of the diffuse astrophysical neutrino flux is however unfortunately not well suited to constraining such effects, further limiting the potential of neutrino decoherence measurements to constrain natural Planck scale physics. 

However, we have seen that should $\gtrsim$TeV neutrinos be detected in association with GRBs, sensitivity to lightcone fluctuations via the arrival time spread of the observed particles (neutrinos, $\gamma$-rays, ...) can likely be significantly enhanced beyond present limits from $\gamma$-ray observations, potentially even far beyond the Planck scale. Current or next generation neutrino telescopes like IceCube, IceCube-Gen2~\cite{Aartsen:2020fgd} and KM3Net~\cite{MARGIOTTA201483} may therefore make crucial contributions in the ongoing search for a quantum theory of gravity. 

\section*{Acknowledgements}

\noindent The authors thank Markus Ahlers for feedback on the paper draft, and Jason Koskinen, Subir Sarkar, Jo\~{a}o Coelho, Mauricio Bustamante, Shashank Shalgar, Mohamed Rameez, Peter Denton and Pilar Coloma for valuable conversations. This work was supported by a Carlsberg Young Researcher Fellowship grant `NuFront: Neutrinos at the Physics Frontier' [case no. CF19-0652] and by VILLUM FONDEN (project no. 13161). The authors would like to acknowledge the contribution of the COST Action CA18108.

%----------------------------------------


\nocite{*}

\bibliography{paper}% Produces the bibliography via BibTeX.

\end{document}